\documentclass[aps,twocolumn,showpacs,superscriptaddress]{revtex4}
\usepackage{amsmath,amssymb}
\usepackage{graphics,graphicx}
\usepackage{dcolumn,bm}
\usepackage{psfrag}
\newcommand{\cu}
{\affiliation{Department of Physics, University of Calcutta,
92 Acharya Prafulla Chandra Road, Kolkata 700009, India.}}

\begin{document}

\title{Minority spin dynamics in non-homogeneous Ising model: diverging timescales and exponents}

\author{Pratik Mullick}%
\cu
\author{Parongama Sen}%
\cu

\normalsize  

\begin{abstract}

 We investigate the dynamical behaviour of the Ising model under a zero temperature quench with the initial fraction of up spins $0\leq x\leq 1$. In one dimension, the known results for persistence probability are verified; it shows algebraic decay for both up and down spins asymptotically with different exponents. It is found that the conventional finite size scaling is valid here. In two dimensions however, the persistence probabilities are no longer algebraic; in particular for $x\leq 0.5$, persistence for the up (minority) spins shows the behaviour $P_{min}(t) \sim t^{-\gamma}\exp(-(t/\tau)^{\delta})$ with time $t$, while for the down (majority) spins, $P_{maj}(t)$ approaches a finite value. We find that the timescale $\tau$ diverges as $(x_c-x)^{- \lambda}$, where $x_c=0.5$ and $\lambda\simeq2.31$. The exponent $\gamma$ varies as $\theta_{2d}+c_0(x_c-x)^{\beta}$ where $\theta_{2d}\simeq0.215$ is very close to the persistence exponent in two dimensions; $\beta\simeq1$. The results in two dimensions can be understood qualitatively by studying the exit probability, which for different system size is found to have the form $E(x) = f\big[(\frac{x-x_c}{x_c})L^{1/\nu}\big]$, with $\nu \approx 1.47$. This result suggests that $\tau \sim L^{\tilde{z}}$, where $\tilde{z} = \frac{\lambda}{\nu} = 1.57 \pm 0.11$ is an exponent not explored earlier.

\end{abstract}


\maketitle

\section{Introduction}

Dynamical phenomena associated with the ordering process of spin systems have been studied since a long time. 
While the domain growth and behaviour of correlation functions were the 
original issues of interest \cite{bray1,pavel}, other phenomena like 
persistence and its variants have been extensively studied more 
recently \cite{persistence1,persist}. 
It is indeed astonishing to note that even after decades of 
research on these topics, newer features have been revealed. For example, 
 in the zero temperature quenching dynamics of Ising model,  
frozen states \cite{lipowski,blanchard} and 
blinkers \cite{frozen1,frozen2,frozen3,blinkers} as nonequilibrium steady states
have been shown to occur  
in dimensions greater than one.  
Recent studies of  the zero temperature quenching dynamics in the two dimensional Ising model
%temperature quenching phenomena 
%in the two dimensional Ising model 
indicated that the domain growth behaviour can be explained in the 
framework of percolation theory \cite{perco1,perco2,perco3}. 
Different ways of employing the zero temperature quench have also been explored in some very recent works in the Ising model \cite{recent1,recent2,recent3}. 
Models with more complicated interactions with binary spin variables have often indicated the existence of dynamical universality classes other than the simple Ising model even in one dimension \cite{binary}.

 Prior to the discovery of the persistence  phenomena
\cite{persistence1} in the Ising and other spin models, 
it was believed that there is only one dynamical exponent which governs 
the  behaviour of the system during the  ordering dynamics. 
The persistence probability showed a power law decay with an exponent 
which could not be related to other known static or dynamic exponents. 
This indicated  that there is a second independent dynamic 
exponent. 
Since the domain size and persistence show power law behaviour, there is essentially no timescale associated with these quantities. 
Physically this means that there are finite probabilities existing 
even at very large timescales. On the other hand, diverging timescales can be associated with   dynamical behaviour close to the critical point.   
The typical relaxation behaviour of the order parameter $m$ as a function of time $t$ is exponential close to the critical temperature: 
$m(t)\sim \exp(-t/\tau)$, with $\tau$ diverging as the critical temperature is approached.

In the present work, we consider the zero temperature single spin flip 
Glauber dynamics of Ising model with nearest neighbour interaction 
when the initial configuration is non-homogeneous, i.e. the fraction of  
up and down spins are different. $x$ is the fraction of up spins ranging 
between $0$ and $1$.  $x=0.5$ is the homogeneous case and due 
to up/down spin symmetry, it is sufficient to take $x \leq 0.5$. 
The spins are otherwise uncorrelated. Three types of  
persistence behaviour is studied: $P_+$ (persistence for up spins), $P_-$ (persistence for down spins) and $P_{total}$ (persistence for the total system). 
The exact expression for the variation of the above mentioned quantities 
are known in one dimension in the thermodynamic limit.
 We have numerically evaluated the persistence probabilities  for 
finite systems 
and  checked that the familiar finite size scaling \cite{fss,fss2} is valid for all values of $x$. 
Our main results are for the two dimensional model, here 
 we have numerically obtained the behaviour of 
$P_+$, $P_-$ and $P_{total}$ and find the presence of a diverging timescale associated with $P_+$ or $P_-$, whichever corresponds to the initial minority.

Another quantity associated with the ordering process that has been studied quite 
intensely in recent years is the   exit probability. 
As a function of $x$ (the initial fraction of up spins), the exit 
probability $E(x)$ is defined as the probability that the final state comprises all up spins. 
 The behaviour of the exit probability helps in understanding qualitatively the persistence probability in the Ising model. Hence we have also conducted a detailed study of the exit probability in two dimensions to gain more insight into the problem.

 In the next two sections, the results for one and two dimensionsal Ising model are discussed respectively. In the last section, summary and some discussions are presented.

\section{Results in one dimension and finite size scaling}

 For systems with non-uniform initial condition i.e. with unequal initial fraction of up and down spins, the exact result for the Ising model in one dimension shows that the density of persistent spins decays algebraically as \cite{prevoter}

\begin{equation}
P_{\pm}(t) \sim t^{{-\theta}_{\pm}}.
\end{equation}
The exponents are different for the up and down spins;

\begin{equation}
{\theta}_{\pm}=\theta(x_\pm)=\frac{2}{\pi^2}\Bigg[\cos^{-1}\Big(\sqrt{2}x_\pm-{\frac{1}{\sqrt{2}}}\Big)\Bigg]^2-\frac{1}{8}
\end{equation}
where $x_+$ $(x_-)$ is the initial concentration of up (down) spins. For equal initial fraction of up and down spins (i.e. for $x_\pm = 0.5$) we have ${\theta}_+ = {\theta}_- = \frac{3}{8}$ \cite{persistence1}. In this paper we follow the notation $x = x_+$ unless otherwise specified.

\par During the coarsening, the typical domain size $D$ shows a power-law growth with time:
\begin{equation}
D(t) \propto t^{1/z},
\label{zeq}
\end{equation}
$z$ being the domain growth exponent. For the Ising model $z = 2$ in all dimensions \cite{bray1}.

\par While the persistence exponents are $x$ dependent, the domain growth exponent $z$ is identical for all $x$. This is easy to explain. Consider, e.g. the growth phenomena for $x = 0.5$. Immediately after the time evolution starts, $x$ will attain a different value. One might as well consider this to be the initial state, hence $z = 2$ for all $x$ \cite{bray1,pavel}.

\par
\begin{figure}
\includegraphics[width=7.5cm]{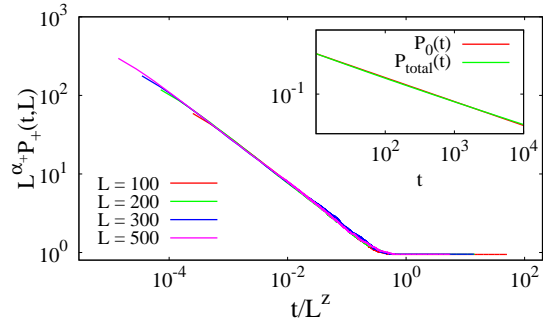}
\caption{Data collapse for persistence probability of up spins for systems with $x=0.3$ in the one dimensional Ising model. This particular collapse was obtained with $\alpha_+ = 1.0$ and $z = 1.95$. Similar data collapse can be obtained for $P_-$ and $P_{total}$ for other values of $x$. Inset shows variation of $P_0(t)$ and $P_{total}(t)$ for $x=0.4$. These calculations were made for a system size $L = 1000$.}
\label{fig1}
\end{figure}

 In one dimension, it is known that for the homogeneous case, the persistence probability obeys finite size scaling. The behaviour of the persistent probability  $P(t,L)$ in a  system with linear dimension  $L$  can be summarised as \cite{fss}
\begin{equation}
P(t,L)=L^{-z\theta}f(t/L^{z})=L^{-\alpha}f(t/L^{z}),
\label{fsseq}
\end{equation}
with $\alpha = z\theta$. We check whether this is also valid when $x \neq 0.5$. Obviously, if valid, the associated exponent $\alpha$ should vary with $x$, since  $z$ is constant. Finite size scaling analysis in one dimension was 
performed for all the three variants of the persistence probability for $x\leq 0.5$. 
Simulations were made for system sizes $L\leq 1000$ taking average over at least 1000 configurations. 
In each Monte Carlo step, $L$ spins are chosen randomly and updated. Asynchronous
updating rule is used and periodic boundary condition imposed. We find that indeed the finite size scaling (\ref{fsseq}) is valid, as the curves for different values of $L$ collapsed with proper choice of $\alpha$, keeping $z \simeq 2$ (Fig. \ref{fig1}). It is checked that these values of $\alpha$ are consistent with the relation $\alpha = z \theta$. In Table \ref{table} we have summarised the values of $\theta$ and $\alpha$ obtained for two different values of $x$. The values of $\theta_+$ and $\theta_-$ differ from the exact values by less than 0.05\%. As for the up, down and total spins $\theta$ is different, 
the corresponding values $\alpha_+$,  $\alpha_- $  and $\alpha _{0}$  
 also turn out to be different.

\par $P_{total}(t)$, the total persistence is expected to be a weighted average of $P_+$ and $P_-$. We calculate the quantity

\begin{equation}
P_0(t)=x_{+}P_{+}(t)+x_{-}P_{-}(t)
\label{totals}
\end{equation}
to check whether $P_0(t)$ and $P_{total}(t)$ are equal and find a very good agreement indeed (inset of Fig. \ref{fig1}).
Even though (\ref{totals}) is valid, it is interesting to note that the exponent $\theta_0$ associated with the 
total persistence probability is different from the $\rm{min}[\theta_+,\theta_-]$, albeit close to it (see Table \ref{table}). 

\begingroup
\begin{table}
\begin{center}
\caption{Persistence and other exponents obtained for Ising model in one dimension using numerical simulation. }
\label{table}
\begin{tabular}{|c||c|c|c|c|c|c|}
\hline
{$x$}&{$\theta_0$}&{$\theta_+$}&{$\theta_-$}&{$\alpha_0$}&{$\alpha_+$}&{$\alpha_-$}\\
\hline
0.3&0.227&0.562&0.212&0.42&1.0&0.40\\
\hline
0.4&0.318&0.463&0.285&0.64&0.94&0.57\\
\hline
\end{tabular}
\end{center}
\end{table}
\endgroup

\par Qualitatively, the algebraic decay of $P_+$ and $P_-$ can be explained from the behaviour of the exit probability $E(x)$ which in the one dimensional Ising model is given by $E(x) = x$. This indicates that there is a finite probability of ending up with either all up or all down spins even for values of $x \neq 0.5$ (except for $x=0$  and $1$).

\section{Results in two dimensions}

\begin{figure}
\includegraphics[width=8.5cm]{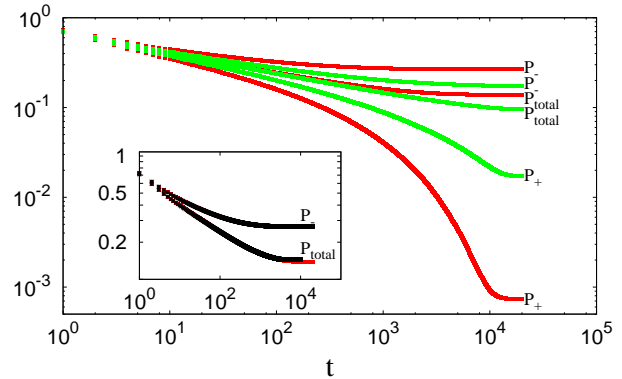}
 \caption{Behaviour of three types of persistence probabilities for two dimensionsal Ising model for two different initial frcation of up spins with $L = 256$. The darker (red) curves correspond to $x = 0.482$ and the lighter (green) ones to $x = 0.492$. Inset shows persistence probabilities for total spins and down spins for $x = 0.482$ for two different system sizes. The lighter (red) curves correspond to $L = 256$ and the darker (black) ones to $L = 128$.}
\label{fig3}
\end{figure}

 In the ordering dynamics of the two dimensional Ising model, the three variants of persistence were studied for various values of $x$ less than $0.5$ on $L \times L$ square lattices with $L\leq 256$. Simulations were performed for 1000 configurations. 
Here also one Monte Carlo time step comprises $L^2$ updates and random asynchronous updating 
rule is used. Helical  boundary condition has been imposed for the simulation in two dimesnions.  In Fig.  \ref{fig3}, the results for $x$ close to 0.5 are shown. 
As $x$ deviates from 0.5, $P_+$ shows a faster decay while both $P_-$ and $P_{total}$ saturate at higher values. 
No appreciable finite size dependence is found for these two persitence probablilities (inset of Fig. \ref{fig3}).
%Finite size dependence is observed in $P_+$ only (inset of Fig. \ref{fig3}), as $L$ increases, $P_+ \rightarrow 0$. 

 By the present convention, $P_-$ is the persistence probability for the down spins, which is the initial majority. Generalising the notation for initial majority and minority spins, we have $P_{maj}$ saturating to a finite value and $P_{min}$ going to zero. To explain why $P_{maj}$ does not show system size dependence, we can argue like this: had it been decreasing with size, it would ultimately go to zero. On the other hand, had it been increasing for larger sizes, it would approach 1. Both the possibilities are unrealistic and hence no system size dependence is observed.
Since $P_{total}$ is basically a weighted average of $P_+$ and $P_-$, its nature is dominated by that of $P_-$,  the majority spins, and hence it does not show finite size dependence also.

\begin{figure}
\includegraphics[width=8cm]{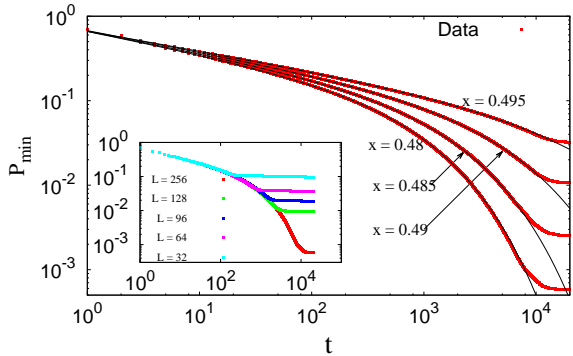}
 \caption{$P_{min}(t)$ against time $t$ for several values of $x$ along with the best fit curves according to eq. (\ref{fit-eq}) are shown. Inset shows persistence probabilities for $x = 0.48$ for five different system sizes. }
\label{fig4}
\end{figure}

 The persistence probabilities do not show algebraic behaviour unlike in the one 
dimensional Ising model. The persistence probability for majority spins as well 
as for the total spins do not show much interesting behaviour as they decay to rather high saturation values and show 
no finite size dependence. 
However in case of the minority spins, finite size effect is apparent and a number of interesting
features are revealed on further analysis. 
 The decay of the persistence probability $P_{min}$ can be fitted to the form 
\begin{equation}
P_{min}(t) \sim t^{-\gamma}\exp(-(t/\tau)^{\delta}),
\label{fit-eq}
\end{equation}
 which is a combination of an algebraic decay accompanied by an stretched exponential cutoff (Fig. \ref{fig4}). 
We find that this fit is valid almost up to the time  where $P_{min}$ saturates to a finite value. 
The fits become more accurate as $x$ approaches 0.5. 
%For finite sizes,   $P_{min}$ does not exactly go to zero and these fits are valid up to a certain value of $t$ which depends on system size. 
%%
%Beyond this value, finite size effects appear such that the probabilities attain a constant value. 
%This is evident from the inset of Fig. ...., where we see that $P_{min}$ represented by $P_{+}$ attains a constant value which decreases with $L$. However, 
%However, the time dependent behaviour,
%before finite size effects
%become visible,  
% is independent of $L$. 
%Similar qualitative behaviour has been  noted at  $x=0.5$ also where, as already mentioned, finite size scaling analysis
%is an useful method of analysis. 
%%the initial regime ($t << L^z$) where no $L$ dependence is present has been shown 
%%to be a purely algebraic function in time in several earlier works. 
%%The associated 
%%exponent is 
% $ \simeq 0.19$ \cite{2disingper,new}.
It is clear that $P_{min}$ extrapolates to a zero value for $L \rightarrow \infty $. However,
the $P_{min}$ versus $t$ curves for different values of $L$ (shown in the inset of Fig. \ref{fig4}) shows that
the finite size dependence is evident only at times beyond which $P_{min}$ saturates.
This is very similar to what happens for $x = 0.5$.
Thus it is sufficient to use eq. (\ref{fit-eq}) for the maximum size ($L = 256$) only.

No exact result is known for the persistence exponent even for $x=0.5$ and one depends only 
on approximate estimates, most of which are numerical \cite{persist,2disingper,new1,new2}. 
%Most recent studies suggest $\theta$ 
%A number of studies for the persistence probability have been made for $x= 0.5$ earlier in two dimensions.  
However, the behavior  of  the persistence probability  has been found to be 
 strictly a power law even for finite system sizes. The
resultant exponent has been shown to have some finite time 
dependence on closer examination \cite{new2}. In the present case
therefore, we have varied the range of time when eq. (\ref{fit-eq}) is used. Noting there is not much variation
in the values of $\tau$ and $\delta$ for different ranges, we have used the average values for 
further analysis.

In order that the eq. (\ref{fit-eq}) is valid for $x \to 0.5$, $\tau$ must diverge at  that limit.   
Also, the  associated  exponent  should be close to $0.20$ according to the 
more recent estimates of the persistence exponent
in two dimensions. 
%Thus 
%  $\tau$ should diverge 
%and $\gamma$ approach a value close to 0.19   
%in the limit $x \to 0.5$. 
To check this, we analysed the behaviour of $\gamma$ and $\tau$ as a function of $(x_c-x)$, where $x_c = 0.5$. 

\begin{figure}
\includegraphics[width=8cm]{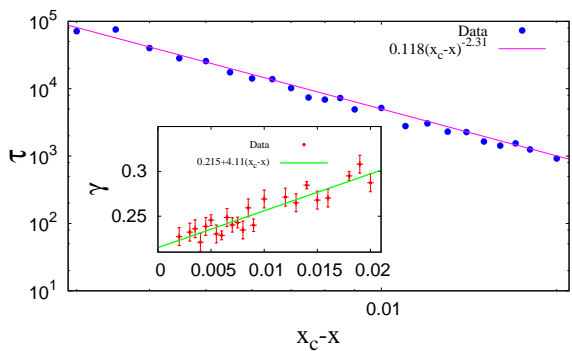}
 \caption{Divergence of $\tau$ as $(x_c-x)^{-\lambda}$ with $\lambda = 2.37$. Inset shows variation of $\gamma$ with $(x_c-x)$ as $\theta_{2d}+c_0(x_c-x)^{\beta}$, with $\theta_{2d} = 0.215$, $c_0 = 2.93$ and $\beta = 1$.}
\label{fig5}
\end{figure}

%These values are obtained by fitting the data with the best fit curve of the form given in eq. (\ref{fit-eq}). 
%However, the fitting range affects the values and we have taken three different ranges by varying the lower limit of
%the range.  In each case, 
The variation of $\tau$ with $(x_c-x)$ is found to be of the form
\begin{equation}
\tau \sim (x_c - x)^{-\lambda},
\label{eq}
\end{equation}
thus showing the divergence at $x \rightarrow 0.5$ (Fig. \ref{fig5}). 
The value of $\lambda$ is numerically equal to 2.31 $\pm 0.16$.  On the other hand, the data for   $\gamma$ 
shows some fluctuations  but apparently has a  linear variation  with $x-x_c$. 
In general   $\gamma$     can be written as  $\theta_{2d} + c$, where 
\begin{equation}
c = c_0 (x_c-x)^{\beta}.
\end{equation}
Putting $\beta = 1$, we obtain $\theta_{2d} = 0.215 \pm 0.004$ which is  indeed very close to the persistence exponent in two dimensions. 
$c_0 =  4.11 \pm 0.37$. The data is shown in Fig. \ref{fig5} (inset). The value of $\theta_{2d}$  is quite insensitive to the value of $\beta$ which  may vary between 
0.8 to 1.2 due to the scatter in the data.
Thus we find the presence of two quantities $\tau$ and $c$ which diverge and vanish respectively as $x \to 0.5$.  
The value of $\delta$ is $\mathcal{O}(10^{-1})$ and it also shows  a slow decrease with $x$. However, it remains finite ($\approx 0.4$) even at values of $x$ very close to 0.5.

\begin{figure}
\includegraphics[width=7.5cm]{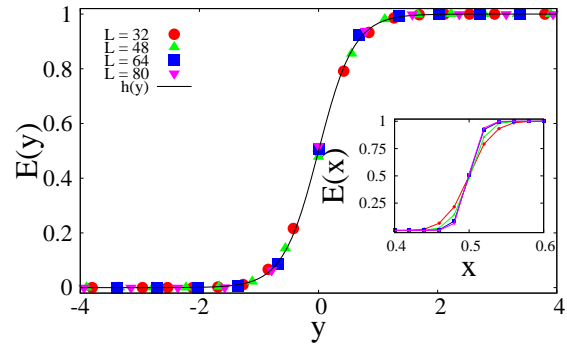}
 \caption{The data collapse for exit probability using $\nu = 1.47$ for different system sizes in two dimensional Ising model. The collapsed plot was fitted according to $h(y)=[\tanh(ay)+1]/2$, where $y = (\frac{x-x_c}{x_c}) L^{1/\nu}$ with $a = 1.67$. Inset shows the unscaled data against initial concentration $x$ of up spins.}
\label{exitfig}
\end{figure}

\par The nature of the persistence probabilities in the two dimensional Ising model can also be explained qualitatively by studying the exit probability. It is known that exactly at $x = 0.5$, the system may not reach the all spins up or down configuration always due to freezing \cite{frozen1,frozen2,frozen3}. Since for finite sizes it is possible that such frozen states may persist at small deviation from $x = 0.5$ as well, we considered only those configurations which led to the true ground states while calculating $E(x)$. $E(x)$ for the two dimensional Ising model shows strong finite size effect and the data indicate that there is a step function behaviour in the thermodynamic limit. Finite size scaling analysis of $E(x)$ was done using the scaling form valid for such a behaviour \cite{exit2}
\begin{equation}
E(x,L)=h\Bigg({\frac{(x-x_c)}{x_c}}L^{1/\nu}\Bigg),
\label{eq9}
\end{equation}
where $h(y)\rightarrow0$ for $y<<0$ and $h(y) = 1$ for $y>>0$ (i.e. a step function like behaviour) shown in Fig. \ref{exitfig}. This scaling argument indicates that $L^{-1/\nu}$ is basically the factor by which the width of the region, where $E(x)$ is not equal to 0 or 1, decreases. The value of $\nu$ from our data collapse is estimated to be $ 1.47 \pm 0.05$. The scaling function $h$ is found to fit with the form
\begin{equation} 
h(y)=[\tanh(ay)+1]/2,
\label{tanh}
\end{equation}
with $a \simeq 1.67$.

\par Since $E(x)$ shows a step function like behaviour in the thermodynamic limit, for an initial concentration of up spins less than 0.5, the probability to reach a configuration with all spins up is simply zero. Hence all the minority spins eventually flip state and therefore $P_{min}(t\rightarrow\infty) = 0$.

\par As already noted, the scaling form (\ref{eq9}) suggests $|x-x_c|$ scales as $L^{-1/\nu}$ and using this in (\ref{eq}) we get $\tau \sim L^{\tilde{z}}$ where $\tilde{z} = \frac{\lambda}{\nu}$. Hence $\tilde{z} = 1.57 \pm 0.11$ can be interpreted as a dynamic exponent connecting time and length scales. The only other known dynamic exponent is $z = 2$ (eq. \ref{zeq}) associated with the domain growth phenomena and is clearly different from $\tilde{z}$. 
%This once again confirms that domain growth and persistence behaviour are independent dynamic phenomena.

\par The scaling function $h$ in (\ref{tanh}) has the same form as found for a class of models with dynamical rules quite different from the Ising model \cite{exitfit}. The value of $\nu$ however is completely different.

\section{Summary and discussion}

\par In summary, we obtained the persistence probability of up, down and total spins for Ising spin-systems using the zero temperature Glauber dynamics in both one and two dimensions. In the initial state, the up spin density $x$ varies between 0 and 1, while the spins are otherwise uncorrelated. In one dimension, the exact results for the persistence exponents are known and we find that the well known finite size scaling form is valid even in the non-homogeneous case.

\par In two dimensions, the results differ drastically, no algebraic decay is observed for the three types of persistence probabilities. The persistence probability $P_{min}$ corresponding to the initial minority spin vanishes. $P_{maj}$ 
saturates to a finite value consistent with the behaviour of the exit probability. The most significant finding  is the diverging time scale associated with the minority spin persistence probability. This timescale is related to the system size through an exponent $\tilde{z} = 1.57 \pm 0.11$, not explored so far to the best of our knowledge. In addition we obtain two other exponents $\beta \simeq 1$ and $\nu \simeq 1.47$.
The values of the exponents obtained in the present study are all close to multiples of 0.5 (within error bars) which suggests that these may be related to the growth exponent $z=2$. 
If this can be shown directly it will lead to a very important and striking result that domain growth and persistence 
probability are no longer independent for $x = 0.5$. However, this a difficult proposition as exact analytical estimate 
of persistence has been possible in one dimension only. Hence the existence of yet another independent
dynamical exponent in the ordering process in the two dimensional Ising model remains  an open question  as of now.

\par Comment: The results presented in this paper have been later improved by generating new data 
especially closer to x = 0.5. A re-analysis of the scaling collapse leads to the value $1.24 \pm 0.05$
for $\nu$. The estimate of $\tilde{z}$ using this value is $\simeq$ 1.86. Hence it can not be strongly stated that $\tilde{z}$
is different from 2.

\par Acknowledgement: The authors thank CSIR (Sanction No. 03(1287)/13/EMR II) for their financial support. Discussion with S. Biswas and P. Ray are also acknowledged.

\end{document}